\documentclass[conference]{IEEEtran}

\usepackage{fancyhdr} 
\usepackage{kantlipsum} 
 \fancypagestyle{plain}{ 
 \fancyhf{} 
 \fancyhead[C]{Conference on \LaTeX} 
  
 } 
 \usepackage{eso-pic}

\usepackage{setspace}
\usepackage{ifpdf}

\usepackage{cite}

\ifCLASSINFOpdf
\else
\fi
\usepackage{graphicx}

\usepackage{amsmath}
\usepackage{algorithmic}

\usepackage{array}

\usepackage{booktabs}
\usepackage{graphicx}

\ifCLASSOPTIONcompsoc
\else
\fi
\usepackage{url}

\makeatletter
\def\@IEEEpubidpullup{2\baselineskip}
\makeatother

\begin{document}
 
\AddToShipoutPictureBG*{ 
 \AtPageUpperLeft{ 
 \setlength\unitlength{1in} 
 \hspace*{\dimexpr0.5\paperwidth\relax}
 \makebox(0,-0.75)[c]{\textbf{2018 IEEE/ACM International Conference on Advances in Social Networks Analysis and Mining (ASONAM)}}}}

%
\title{TED Talk Recommender Using Speech Transcripts}



%
\author{\IEEEauthorblockN{Jaehoon Oh\IEEEauthorrefmark{1}\IEEEauthorrefmark{2},
Injung Lee\IEEEauthorrefmark{1}\IEEEauthorrefmark{2},
Yeon Seonwoo\IEEEauthorrefmark{1}\IEEEauthorrefmark{2}, 
Simin Sung\IEEEauthorrefmark{1},
Ilbong Kwon\IEEEauthorrefmark{1} and
Jae-Gil Lee\IEEEauthorrefmark{1}\IEEEauthorrefmark{3}}
\IEEEauthorblockA{\IEEEauthorrefmark{1}Korea Advanced Institute of Science and Technology (KAIST), Daejeon, South Korea\\\IEEEauthorrefmark{2}These authors contributed equally.\\\IEEEauthorrefmark{3}Jae-Gil Lee is the corresponding author.\\Email: \{jhoon.oh, edndn, yeon.seonwoo, sa7377, hellomacro77, jaegil\}@kaist.ac.kr}}


\maketitle

\IEEEoverridecommandlockouts 
 \IEEEpubid{\parbox{\columnwidth}{\vspace{8pt} 
 \makebox[\columnwidth][t]{
 \textbf{IEEE/ACM ASONAM 2018, August 28-31, 2018, Barcelona, Spain}} 
 \makebox[\columnwidth][t]{
 \textbf{978-1-5386-6051-5/18/\$31.00~\copyright\space2018 IEEE}} \hfill} \hspace{\columnsep}\makebox[\columnwidth]{}} 
 \IEEEpubidadjcol 

\begin{abstract}
Nowadays, online video platforms mostly recommend related videos by analyzing user-driven data such as viewing patterns, rather than the content of the videos. However, content is more important than any other element when videos aim to deliver knowledge. Therefore, we have developed a web application which recommends related TED lecture videos to the users, considering the content of the videos from the transcripts. \textit{TED Talk Recommender} constructs a network for recommending videos that are similar content-wise and providing a user interface. Our demo system is available at http://dmserver6.kaist.ac.kr:24673/.
\end{abstract}


%
\IEEEpeerreviewmaketitle

\section{Introduction}
A vast number of videos are being created and uploaded to video-sharing websites such as YouTube. To assist users in finding videos they want to watch, recommendation services are being developed\cite{cheng2007understanding}. For instance, Davidson et al. developed a system that recommends videos in relation to users’ watching patterns\cite{davidson2010youtube}. In other words, most of the recommendation systems for online videos focused on in-formation given by users rather than the content of the videos. However, the content of videos is especially important when they have the purpose of delivering information or knowledge, such as in TED.com. With technical improvements for speech recognition, transcripts of videos are automatically generated and serviced\cite{hinton2012deep}. Also, more precise semantic understanding has been enabled by applying deep learning methods\cite{shen2014learning}. Therefore, we developed an application that suggests TED talks videos according to the similarities among the content of the videos. Our web application analyzes transcripts, retrieves contents of the videos using \textit{Doc2vec} in the \textit{Gensim} package\cite{le2014distributed}, calculates similarities between these, and recommends semantically similar ones to the users.

\section{Implementation}
We obtained the TED talk datasets from Kaggle\cite{kaggleurl} including data on 2,400 TED talks with title, speaker, tagging, transcript, and so on. From the datasets, we mainly used transcripts that have on average 3,000 words.
\begin{description}
\item[$\bullet$ \textbf{Step}]\textbf{1:} We derived positive and negative scores from the transcripts using language assessment by Mechanical Turk (labMT)\cite{peter2011temporal}. A higher score represents more positive content, and signifies that the video’s content has a positive emotional influence on the audience
\item[$\bullet$ \textbf{Step}]\textbf{2:} We applied TF-IDF analysis to determine which words represent the semantics of the TED talk. The words with higher TF-IDF values were used to form word-clouds that characterize each video.
\item[$\bullet$ \textbf{Step}]\textbf{3:} We applied \textit{Doc2vec} of the \textit{Gensim} package to derive vectors of transcripts. The vector representations are trained and we used them to calculate cosine similarity among those document embeddings. We used vector dimensionality 200 and context size 8 as hyper-parameters.
\end{description}

\section{System Demonstration}
The back-end of our web application analyses such as calculation of similarities, community detection, and sentiment analysis were conducted using Python. Regarding the user interface of our web application, the networks of similar videos were constructed using d3.js.

Fig. 1 is the home page of our web application, and the left panel shows the list of titles of the 2,400 TED talk videos. The center panel shows the main network where videos are represented as nodes and their similarities as edges. The main network displays only the top 1 percent of similarity-scored relationships as the edges. The color of nodes indicates the sentiment score of the video: blue for negativity and red for positivity. The size of the node signifies the number of views of each video, and the nodes are grouped depending on communities detected. 

\begin{figure}[ht]
\center
\includegraphics[width=0.45\textwidth]{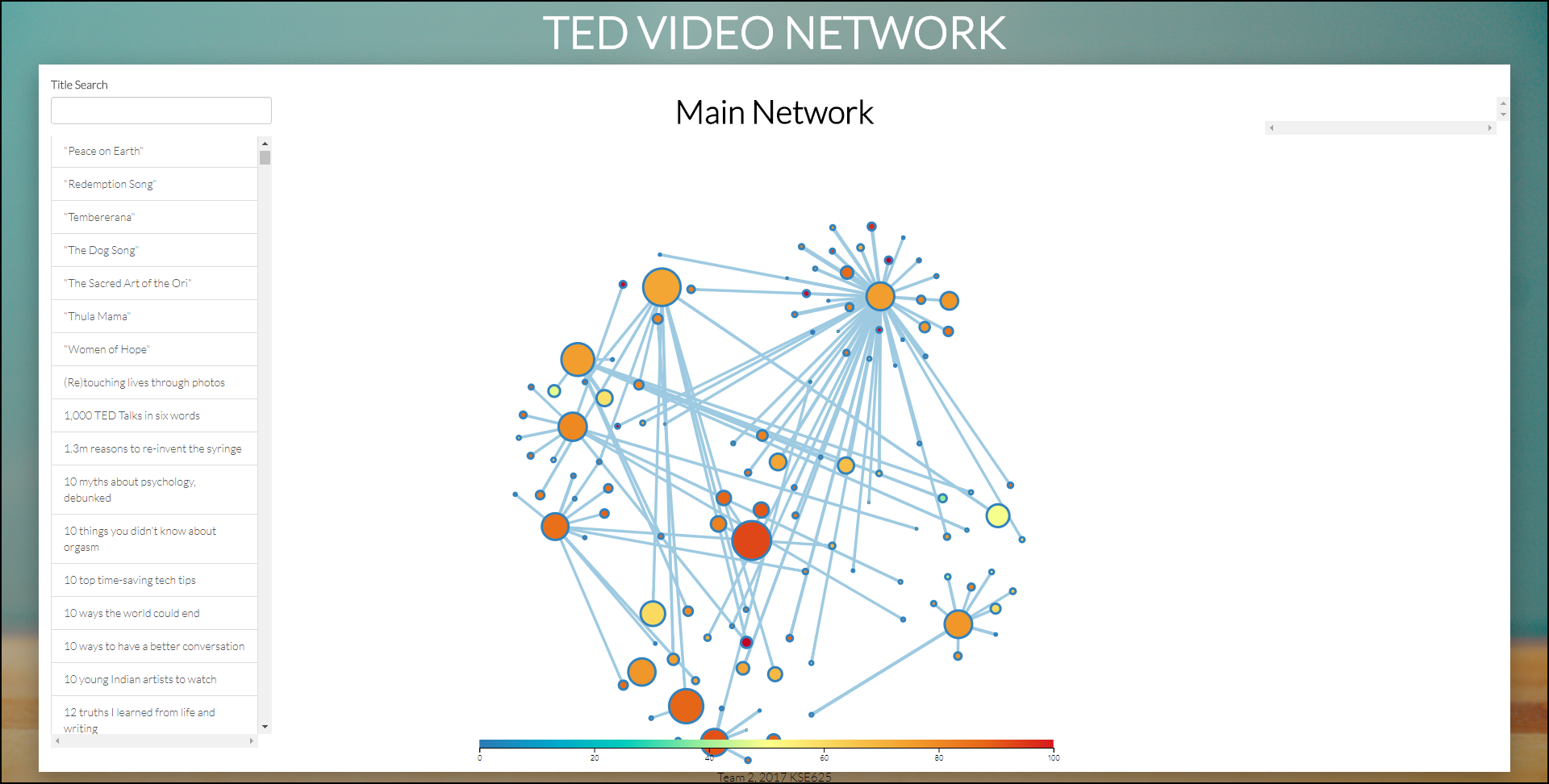}
\caption{The main network of the TED talks recommender}
\end{figure}

When a mouse pointer hovers over a node, the title of the video appears, and the right panel shows information about that node: a word-cloud that summarizes the contents of the video, and the list of other videos that share similar contents (Fig. 2). By hovering the mouse pointer over the nodes, users can surf around TED talk videos.

\begin{figure}[ht]
\center
\includegraphics[width=0.45\textwidth]{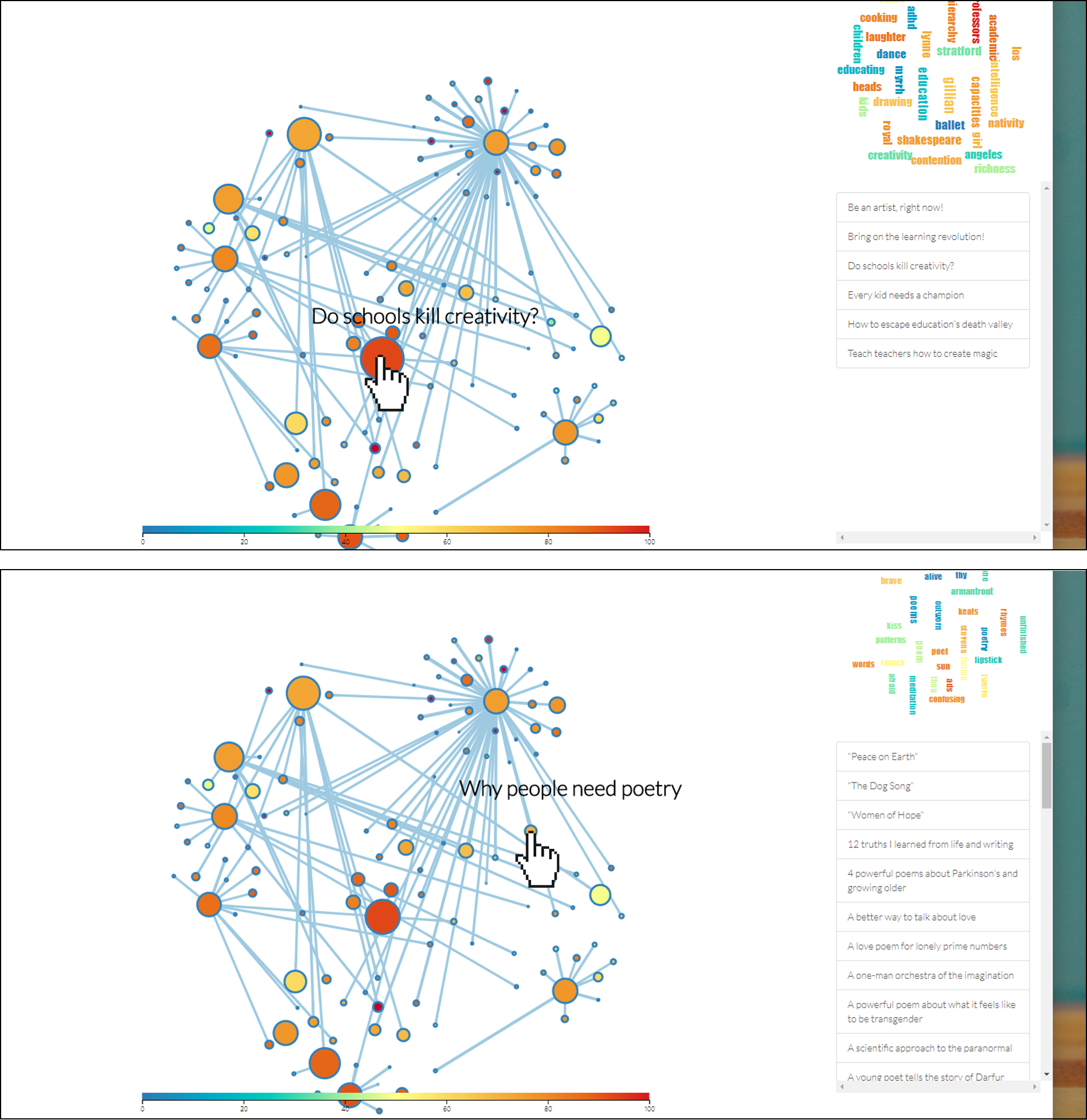}
\caption{Interaction with the visualized network. Hovering a mouse pointer over a node shows the title, word cloud, and similar ones of the video.}
\end{figure}

To be directed to recommended videos, similar to those previously viewed, the user can click on the video title displayed on the left panel, or search by typing its title into the search box. Fig.3 is the screen displayed when a user searched for the video titled ``3 ways the brain creates meaning." The neighbor network of this video appears on the central panel. It shows highly recommended videos that share similar content. By hovering the mouse over the nodes, the right panel displays a word-cloud and a list of relevant videos from the most related to the least. By clicking on one of the titles listed, the user lands on that video’s web-page on TED.com.

\begin{figure}[ht]
\center
\includegraphics[width=0.45\textwidth]{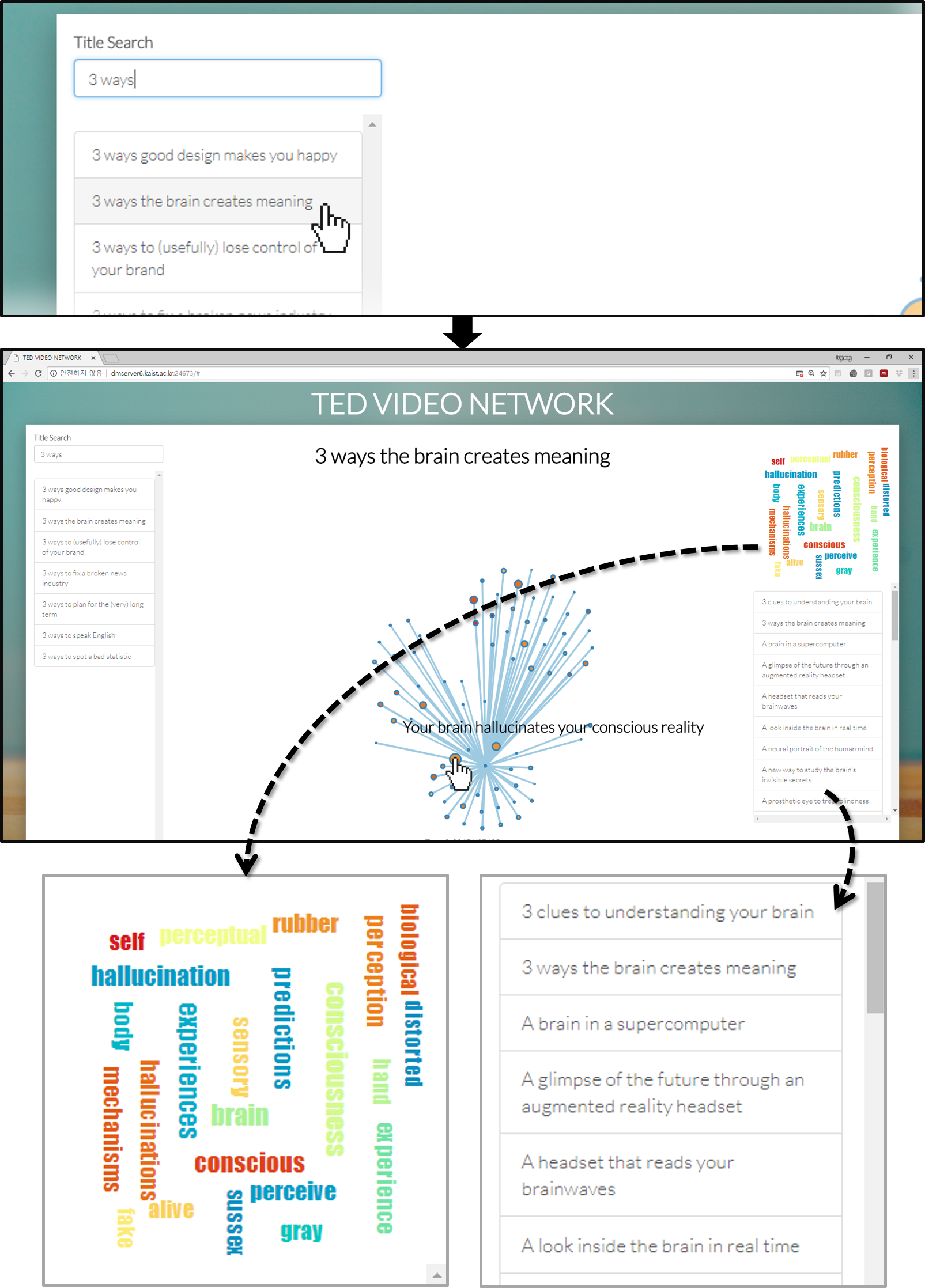}
\caption{Example of a retrieved video and its detailed result}
\end{figure}

Table 1 compares the videos recommended by TED.com and those by our method, for a lecture titled ``Do schools kill creativity?" TED.com provides a list of up to six related videos per TED talk, while our method recommends more than ten videos from the most related to the least. An average of two videos resulted as the intersection of two sets, a set of recommended videos from TED.com and a set using our method (Part B). Those intersecting videos had mostly the same speaker. Some videos originally suggested by TED.com video curators were not included in the results of our recommender (Part A). By looking through the data set from Kaggle, we were able to ascertain that those videos shared common tags. Some results, including lecture \textit{d} which might be less relevant than the lecture \textit{g}, showed the limitations of such \textit{tag-oriented} recommendations. The main subject of the selected TED talk and lecture \textit{g} emphasize ``Children have their own creativity, and teachers should try to keep their children creative," but that of lecture \textit{d} emphasizes, ``Children have a right to education without discrimination." Part C is about the related videos that were not listed in related videos data from TED.com but were recommended by our method.

\begin{table}[]
\centering
\caption{Comparison of videos recommended by \ TED.com and our application}
\label{my-label}
\resizebox{\columnwidth}{!}{%
\begin{tabular}{@{}ll@{}}
\toprule
 & \multicolumn{1}{c}{\textbf{Videos recommended only by TED.com (Part A)}} \\ \midrule
\begin{tabular}[c]{@{}l@{}}a\\ b\\ c\\ d\end{tabular} & \begin{tabular}[c]{@{}l@{}}``How to fix a broken school? Lead fearlessly, love hard" - Linda Cliatt-Wayman, \\ ``Education innovation in the slums" - Charles Leadbeater, \\ ``A short intro to the Studio School" - Geoff Mulgan, \\ ``How America's public schools keep kids in poverty" - Kandice Sumnet\end{tabular} \\ \midrule
 & \multicolumn{1}{c}{\textbf{Videos recommended by both TED.com and our application (Part B)}} \\ \midrule
\begin{tabular}[c]{@{}l@{}}e\\ f\end{tabular} & \begin{tabular}[c]{@{}l@{}}``Bring on the learning revolution!" - Ken Robinson,\\ ``How to escape education's death valley" - Ken Robinson\end{tabular} \\ \midrule
 & \multicolumn{1}{c}{\textbf{Videos recommended only by our application (Part C)}} \\ \midrule
\begin{tabular}[c]{@{}l@{}}g\\ h\\ i\\ j\\ k\\ l\\ m\\ n\end{tabular} & \begin{tabular}[c]{@{}l@{}}``Every kid needs a champion" - Rita Pierson, \\ ``Be an artist, right now!" - Young-ha Kim, \\ ``Teach teachers how to create magic" - Christopher Emdin, \\ ``4 pillars of college succcess in science" - Freeman Hrabowski, \\ ``What we think we know" - Jonathan Drori, \\ ``Your brain on improv" - Charles Limb,\\ ``A on-woman global village" - Sarah Jones,\\ ``The transformative power of classical music" - Benjamin Zander\end{tabular} \\ \bottomrule
\end{tabular}%
}
\end{table}

\section{Conclusion}
This paper has introduced the new concept of a recommendation system for TED talk videos and has developed the idea into a web application. Our application recommends semantically related videos by measuring similarities of video transcripts, using deep learning techniques. Compared to tag-based recommendation, our method has the possibility to provide better-related videos, in terms of both quality and quantity, in considering that most lectures in Part C look very relevant. Though this research we could confirm that our transcript-based recommendation method is applicable to videos that have speeches.

Moreover, with the improvement of speech recognition technology, our method is expected to be expanded to other video platforms in addition to TED.com. In addition to the video platforms such as YouTube, social network services that provide video sharing can also benefit from our method. By applying it to user-created videos, the contents of those videos can be analyzed and used to build the semantic networks of the videos. As a result, the users will be able to easily find the videos relevant to their interests, without being misled by inappropriate recommendations.

\section{Acknowledgement}
This research, ``Geospatial Big Data Management, Analysis and Service Platform Technology Development," was supported by the MOLIT(The Ministry of Land, Infrastructure and Transport), Korea, under the national spatial information research program supervised by the KAIA(Korea Agency for Infrastructure Technology Advancement) (18NSIP-B081011-05). 

\newpage

\bibliographystyle{IEEEtran}
\bibliography{references}

\end{document}